\documentclass[preprint,showpacs,preprintnumbers,amsmath,amssymb,aip]{revtex4-1}

\usepackage{graphicx}
\usepackage{dcolumn}
\usepackage{bm}
\usepackage{psfrag}
\usepackage[T1]{fontenc}

\begin{document}

\preprint{APS/123-QED}

\title{Thermal convection in compressible gas with spanwise rotation}

\author{K. L\"udemann and A. Tilgner}

\affiliation{Institute of Astrophysics and Geophysics, University of G\"ottingen,
Friedrich-Hund-Platz 1, 37077 G\"ottingen, Germany }

\date{\today}

\begin{abstract}
We simulate numerically convection in a rectangular cell filled with an ideal gas
rotating about an axis perpendicular to the direction of gravity. This
configuration corresponds to an experiment with a convection cell placed in a
rapidly rotating centrifuge in which the centrifugal force plays the role of
gravity. The compressibility of the gas in combination with the rotation of the
cell leads to a drifting mode at the onset of convection. The drift persists
despite the presence of sidewalls and is rapid enough to cause the anelastic
approximation to fail at parameters typical of realizable laboratory
experiments. The global spanwise rotation forces the flow to be 2D unless the
Rayleigh number is sufficiently large. The main properties of compressible
convection in 2D and 3D are compared.
\end{abstract}

\maketitle

\section{Introduction}
There is an ongoing effort to study Rayleigh-B\'enard convection experimentally at 
large Rayleigh numbers. Experiments routinely use fluids with low thermal
diffusivity and viscosity and possibly a cryogenic environment to realize
extreme Rayleigh numbers \cite{Ahlers09}. More recently, two experiments employed convection
cells mounted on a centrifuge to create a large effective gravity. One of them
operated with water as a working fluid \citep{Jiang20} while the other filled
the cell with a gas so that effects of compressibility became accessible to
experimental study \citep{Menaut19}. Convection in a centrifuged cell occurs in
a rotating frame of reference in which centrifugal and Coriolis forces appear.
The centrifugal force is desired in this context. The effective gravity created
by the centrifuge is approximately a uniform force field
if the convection cell is for example the gap
between two cylinders coaxial with the rotation axis, the inner cylinder cooled
and the outer cylinder heated, and if the gap is small compared with the radius
of curvature of the inner cylinder. \citet{Jiang20} used a cylindrical cell,
whereas \citet{Menaut19} used a rectangular cell with the rotation axis outside
the cell and parallel to one of the sides. The Coriolis force is a nuisance if 
the only intention of
the centrifuge is to generate a large gravity, but it is part of the model if
the centrifuged cell is supposed to emulate convection in the equatorial region
of a rotating spherical self gravitating body. Since the Coriolis acceleration is
proportional to velocity, its magnitude may be large or small compared with the
advection term in the momentum equation, so that the relevance of the Coriolis
force depends on the flow state. The most important effect of the Coriolis term
in incompressible fluids is to favor 2D structure. The Coriolis force has zero curl 
and can be balanced by a pressure gradient as long as the velocity
field is two dimensional and independent of a coordinate directed along the axis
of rotation. One can therefore conclude that convection in a centrifuged cell
will behave like 2D convection if the Coriolis force is a dominating
force in the momentum equation. The criterion governing the transition from 2D
to 3D convection flows with spanwise rotation was studied by \citet{Luedem22}.

Compressibility adds a new effect. Assume that convection is nearly 2D and
independent of the coordinate along the rotation axis, and
that density is stratified in a direction perpendicular to the rotation axis. As
a vortex aligned with the rotation axis moves towards a region of different
density, its moment of inertia varies and its vorticity has to change as well to
conserve angular momentum. This vorticity generation mechanism is commonly
called \cite{Evonuk08,Glatzm09,Verhoe14} ''compressional $\beta$-effect'' in
analogy with the ''topographic $\beta$-effect''. In the well studied topographic
$\beta$-effect, vorticity is generated by vortices moving in an incompressible
fluid from deep to shallow layers, so that the height and therefore the radius
of the vortices vary, and hence their moment of inertia. Compressional and
topographic $\beta$-effects have different physical origins, but their effect on
2D flows is equivalent. Convection in a centrifuged cell filled with
incompressible fluid is insensitive to any $\beta$-effect if the height of the
cell measured along the rotation axis is everywhere the same, as for example
in the rectangular cell used by \citet{Menaut19}. However, a
compressible fluid in the same cell will always experience the compressional
$\beta$-effect to some degree. 

This paper is concerned with the behavior of compressible ideal gas in a
centrifuged convection cell and is inspired by the experiments by
\citet{Menaut19}. The choice of the control parameters in the simulations is
guided by the experimental application. The density variations in the
simulations are similar to, or up to an order of magnitude larger than what can
be realized in experiments, and they are much smaller than in typical 
astrophysical applications. The density in the Earth's core, however, varies by
20-30\%, which is in the range of experimentally accessible density variations.
The rotation rates necessary to experimentally obtain density variations of this
magnitude are so large that the resulting flows are mostly 2D. This is why the
simulations will pay particular attention to 2D flows.

The paper is organized so that it covers sequentially the different types of
flow that one may encounter when varying the Rayleigh number at fixed Ekman
number. The crucial effect at low Rayleigh number is the onset of convection. 
The compressional $\beta$-effect delays the onset of convection as will be
computed in section \ref{section:onset}. In the opposite limit of very large
Rayleigh number, the convection flow is fast and the advection term overwhelms
the Coriolis term in the momentum equation so that a fully 3D flow must be
expected. Such a flow will turn into 2D convection if the Rayleigh number is
lowered at constant Ekman number. The question is whether the distinction
between 3D and 2D matters for experiments which cannot visualize the full flow
field and which can only measure local or characteristic velocities, or a global
quantity like the total heat transport across the cell. Two and three dimensional convection
will be compared in section \ref{section:compare}, and it will be seen that
there are detectable differences between 2D and 3D convection. This immediately
prompts the next question, which is whether there is a transition within the 2D
flows. The compressional $\beta$-effect was crucial for the onset of convection
in section \ref{section:onset}, but we expect it to become less and less relevant
if the Rayleigh number is increased and rotational effects altogether become
less efficient. Section \ref{section:transition} locates the transition from 2D
flows in which the compressional $\beta$-effect is important to 2D flows in
which the compressional $\beta$-effect is irrelevant.
Finally, section \ref{section:mean_flow} deals with the mean
flow that may appear in rotating convection.

\section{The mathematical model}
\label{section:model}

We consider in a Cartesian coordinate system $x,y,z$ a rectangular cell of size
$d$ along $z$ and size $L$ along $x$ and $y$. The cell rotates at rate $\Omega$
about an axis parallel to the $y-$axis. In the experiments, the rotation is
responsible both for the Coriolis and the centrifugal acceleration, but the
latter also depends on the distance from the cell to the rotation axis. The
effective gravity therefore is an independent parameter which we call $g$. This
gravitational force is assumed to act along the negative $z-$direction. The cell is filled
with ideal gas which is characterized by constant specific heat capacities at fixed
volume and pressure, $C_V$ and $C_p$. The dynamic viscosity $\mu$ and heat
conductivity $k$ are also assumed constant. The kinematic viscosity $\nu$ and
the thermal diffusivity $\kappa$ therefore depend on the density $\rho$ as
$\nu=\mu/\rho$ and $\kappa=k/(\rho C_p)$. The two boundaries perpendicular to
$z$ are assumed to be at fixed temperatures. The colder of the two boundaries is
nearer to the rotation axis in the experiment, but in the astrophysical or
geophysical application, this would be the outer boundary. The temperature of the
cold boundary will be denoted by $T_o$, where the subscript stands for
''outer''. This notation is slightly confusing in the experimental context, but
we keep it for easier comparison with earlier simulations of 3D nonrotating
compressible convection \cite{Tilgne11}. Similarly, $\rho_o$, $\nu_o$ and
$\kappa_o$ denote $\rho$, $\nu$ and $\kappa$ evaluated at the cold boundary in
the conductive state in which the gas is at rest. The temperature difference
applied to the cell is  $\Delta T$.

The temperature at any position $\bm r$ and time $t$ will be denoted by $T(\bm
r,t)+T_o$. The temperature variable $T(\bm r,t)$, the pressure $p(\bm r,t)$ and
the velocity $\bm v(\bm r,t)$ are then governed by the following equations:
\begin{eqnarray}
\partial_t \rho + \nabla \cdot (\rho \bm v) 
&=& 0
\label{eq:conti_dim} \\
\rho [\partial_t \bm v +(\bm v \cdot \nabla) \bm v] + 2 \Omega \bm{\hat y} \times \bm v
&=& 
-\nabla p +\rho \bm g +
\mu \left[\nabla^2 \bm v + \frac{1}{3} \nabla (\nabla \cdot \bm v)\right]
\label{eq:NS_dim} \\
\partial_t T + \bm v \cdot \nabla T 
&=&
\frac{C_p}{C_V}\kappa \nabla^2 T -
\frac{p}{\rho C_V} \nabla \cdot \bm v + \frac{2\mu}{\rho C_V} \left[e_{ij}-\frac{1}{3} (\nabla \cdot
\bm v) \delta_{ij}\right]^2
\label{eq:T_dim} \\
p &=& \rho R (T+T_o)
\label{eq:state} 
\end{eqnarray}
Hats denote unit vectors and summation over repeated indices is implied.
The gas constant $R$ in the equation of state is given by $ R=R_u/m $, with the
molar mass $m$ and the universal gas constant $R_u=8.314 J~ mol^{-1}~ K^{-1}$.
It follows from thermodynamics that $R=C_p-C_V$. The strain rate tensor $e_{ij}$
is given by $e_{ij}=\frac{1}{2}(\partial_j v_i + \partial_i v_j)$.

We will now introduce nondimensional variables. All lengths will be expressed in
multiples of $d$, time and density in multiples of $d^2/\kappa_o$ and $\rho_o$,
and the variable $T$ in multiples of $\Delta T$. Using the same symbols for the
nondimensional variables as for the dimensional variables, one arrives at the
equations of evolution
\begin{equation}
\partial_t \rho + \nabla \cdot (\rho \bm v) = 0
\label{eq:conti}
\end{equation}

\begin{eqnarray}
\partial_t \bm v +(\bm v \cdot \nabla) \bm v + 2 \frac{\mathrm{Pr}}{\mathrm{Ek}}\bm{\hat y} \times \bm v 
& = &
-\frac{1}{\rho}\nabla \left[\left(T+\frac{T_o}{\Delta T}\right)\rho\right]\frac{1}{\gamma}
\frac{H_o}{d} \mathrm{Pr} ~ \mathrm{Ra} 
\label{eq:NS}
\\
& + & \hat{\bm z}\mathrm{Pr} ~ \mathrm{Ra}\frac{T_o}{\Delta T}+
\left[\nabla^2 \bm v + \frac{1}{3} \nabla (\nabla \cdot \bm v)\right]\frac{1}{\rho}\mathrm{Pr}
\nonumber
\end{eqnarray}

\begin{eqnarray}
\partial_t T + \bm v \cdot \nabla T & = & \frac{\gamma}{\rho}\nabla^2 T -
(\gamma-1)\left(T+\frac{T_o}{\Delta T}\right) \nabla \cdot \bm v 
\label{eq:T}
\\ 
& + & \left[e_{ij}-\frac{1}{3} (\nabla \cdot \bm v) \delta_{ij}\right]^2 \frac{1}{\rho} 2 \gamma
(\gamma-1)
\frac{d}{H_o} \frac{1}{\mathrm{Ra}}
\nonumber
\end{eqnarray}
and the boundary conditions
\begin{equation}
T(z=1)=0 ~~~,~~~ T(z=0)=1 ~~~,~~~ \bm v(z=1)= \bm v(z=0) =0.
\label{eq:bc}
\end{equation}
if the cold and warm boundaries are no slip, which will be our standard choice. 
If on the contrary these boundaries
are assumed to be stress free (which will also prove useful below), the conditions on
velocity are replaced with
\begin{equation}
v_z(z=1)=0 ~~~,~~~ v_z(z=0)=0 ~~~,~~~ 
\partial_z v_x(z=1)= \partial_z v_x(z=0) = \partial_z v_y(z=1)= \partial_z
v_y(z=0) = 0.
\end{equation}
Two types of lateral boundary conditions will be considered: either stress free
and thermally insulating sidewalls enclose a square of cross section 
$L \times d$, or periodic boundary conditions are applied at the edges of
this square. Most simulations will use $L/d=2$.

There are several control parameters in the equations of evolution. The Ekman
number $\mathrm{Ek}$,
\begin{equation}
\mathrm{Ek} = \frac{\nu_o}{\Omega d^2}
\end{equation}
the Rayleigh number $\mathrm{Ra}$,
\begin{equation}
\mathrm{Ra} = \frac{g d^3 \Delta T}{T_o \kappa_o \nu_o}
\end{equation}
and the Prandtl number $\mathrm{Pr}$,
\begin{equation}
\mathrm{Pr} = \frac{\nu}{\kappa}.
\end{equation}
The Prandtl number is constant throughout the fluid in the present model and will be set to 0.7 in
all simulations. The adiabatic exponent $\gamma$ is set to its value for a
monoatomic gas
\begin{equation}
\gamma=C_p/C_V=\frac{5}{3}
\end{equation}
in which case it is also justified to ignore bulk viscosity in eqs.
(\ref{eq:conti_dim}-\ref{eq:state}).
The adiabatic scale height for density at the cold boundary, $H_o$, is given by
$H_o=\gamma R T_o / g$ and specifies the density stratification through the
control parameter $d/H_o$.

The static conduction profile is established if $\bm v=0$ and has temperature
and density profiles $T_s(z)$ and $\rho_s(z)$ given by
$T_s(z)=1-z$ and
\begin{equation}
\rho_s = \rho_o \left( \frac{1}{1+\frac{1-z}{T_o/\Delta T}} \right) ^{1-\gamma
\frac{d}{H_o}\frac{T_o}{\Delta T}}
\label{eq:density_init}
\end{equation}
in which the last control parameters $\Delta T/T_o$ and $\rho_o$ appear. In order to recover the
Boussinesq equations in the limit of $\frac{d}{H_o}$ and $\frac{\Delta T}{T_o}$
tending to zero, $\rho_o$ was always set to 1.

To prepare the computations of the linear onset, we introduce the variables
$T_1$ and $\rho_1$ which describe the deviations from the static profiles as
$T=T_s+T_1$ and $\rho=\rho_s+\rho_1$ and note that the static state is in the
hydrostatic equilibrium described by eq. (\ref{eq:NS}) with $\bm v=0$:
\begin{equation}
\nabla \left[\left(T_s + \frac{T_o}{\Delta T}\right) \rho_s\right] = - \hat{\bm z} \rho_s
\gamma \frac{d}{H_o} \frac{T_o}{\Delta T}.
\label{eq:hydrostat}
\end{equation}
Expressing eq. (\ref{eq:NS}) in terms of $T_1$ and $\rho_1$ turns eq.
(\ref{eq:NS})
with the help of eq. (\ref{eq:hydrostat}) into
\begin{eqnarray}
\frac{\rho}{\rho_s} \left[\partial_t \bm v +(\bm v \cdot \nabla) \bm v 
+ 2 \frac{\mathrm{Pr}}{\mathrm{Ek}} \bm{\hat y} \times \bm v\right]
&=&
\left[-\nabla \frac{p_1}{\rho_s} - \frac{p_1}{\rho_s} \nabla \ln \rho_s \right]
\frac{1}{\gamma} \left( \frac{cd}{\kappa} \right)^2 \frac{\Delta T}{T_o}
- \frac{\rho_1}{\rho_s} \hat{\bm z} \mathrm{Pr} \mathrm{Ra} \frac{T_o}{\Delta T}\nonumber\\
&+& \frac{\mathrm{Pr}}{\rho_s} \left[\nabla^2 \bm v + \frac{1}{3} \nabla (\nabla \cdot \bm v)\right]
\end{eqnarray}
with the shorthand notation
\begin{equation}
p_1 = \left( T_s + \frac{T_o}{\Delta T} \right) \rho_1 + T_1 \rho.
\end{equation}
The speed of sound $c$ is given by $c^2=\gamma R T$ so that
\begin{equation}
\frac{H_o}{d} \mathrm{Pr} \mathrm{Ra} = \left( \frac{cd}{\kappa} \right)^2
\frac{\Delta T}{T_o}.
\end{equation}

By using once again eq. (\ref{eq:hydrostat}), one can eliminate $\nabla \ln
\rho_s$ in favor of $\nabla [(T_s + \frac{T_o}{\Delta T})]$ to obtain
\begin{eqnarray}
\frac{\rho}{\rho_s} \left[\partial_t \bm v +(\bm v \cdot \nabla) \bm v
+ 2 \frac{\mathrm{Pr}}{\mathrm{Ek}} \bm{\hat y} \times \bm v\right]
&=&
\left[-\nabla \frac{p_1}{\rho_s} + \frac{p_1}{\rho_s} \nabla \ln \left( T_s +
\frac{T_o}{\Delta T} \right) \right]
\frac{1}{\gamma} \left( \frac{cd}{\kappa} \right)^2 \frac{\Delta T}{T_o}
\nonumber \\
&+& 
\frac{\rho}{\rho_s}
\frac{T_1}{T_s + \frac{T_o}{\Delta T}} \hat{\bm z} \mathrm{Pr} \mathrm{Ra} \frac{T_o}{\Delta T}
+ \frac{\mathrm{Pr}}{\rho_s} \left[\nabla^2 \bm v + \frac{1}{3} \nabla (\nabla \cdot \bm v)\right]
\end{eqnarray}
This equation still is completely equivalent to eq. (\ref{eq:NS}). Linearization
now proceeds by replacing $\rho/\rho_s$ with 1, by
removing the advection term, and by replacing every product $T_1 \rho$ with $T_1
\rho_s$, also in $p_1$, which leads to
\begin{eqnarray}
\partial_t \bm v + 2 \frac{\mathrm{Pr}}{\mathrm{Ek}} \bm{\hat y} \times \bm v 
&=&
\left\{ -\nabla\left[\left( T_s +\frac{T_o}{\Delta T} \right) \frac{\rho_1}{\rho_s} + T_1\right]
-\left[\left( T_s +\frac{T_o}{\Delta T} \right) \frac{\rho_1}{\rho_s} + T_1\right] 
\frac{\hat{\bm z}}{T_s + \frac{T_o}{\Delta T}} \right\}
\frac{1}{\gamma} \left( \frac{cd}{\kappa} \right)^2 \frac{\Delta T}{T_o}
\nonumber \\
&+& 
\frac{T_1}{T_s + \frac{T_o}{\Delta T}} \hat{\bm z} \mathrm{Pr} \mathrm{Ra} \frac{T_o}{\Delta T}
+ \frac{\mathrm{Pr}}{\rho_s} \left[\nabla^2 \bm v + \frac{1}{3} \nabla (\nabla \cdot
\bm v)\right].
\end{eqnarray}

In the experimental context, we will be especially interested in the case of
small $\frac{\Delta T}{T_o}$, so that 
$T_s+\frac{T_o}{\Delta T} \approx \frac{T_o}{\Delta T}$ since $|T_s| \leq 1$,
and of small $\frac{d}{H_o}$ so that $\rho_s \approx \rho_o$. The linearized
momentum equation then becomes
\begin{eqnarray}
\partial_t \bm v + 2 \frac{\mathrm{Pr}}{\mathrm{Ek}} \bm{\hat y} \times \bm v 
&=&
\left\{ -\nabla\left[ \frac{\rho_1}{\rho_o} + T_1 \frac{\Delta T}{T_o} \right]
- \hat{\bm z} \frac{\Delta T}{T_o} \left[ \frac{\rho_1}{\rho_o} + T_1 \frac{\Delta T}{T_o} \right]
\right\}
\frac{1}{\gamma} \left( \frac{cd}{\kappa} \right)^2 
\nonumber \\
&+& 
T_1 \hat{\bm z} \mathrm{Pr} \mathrm{Ra} 
+ \frac{\mathrm{Pr}}{\rho_o} \left[\nabla^2 \bm v + \frac{1}{3} \nabla (\nabla \cdot
\bm v) \right].
\label{eq:lin_mom}
\end{eqnarray}
The linearized temperature equation reads
\begin{equation}
\partial_t T_1 - v_z = \frac{\gamma}{\rho_s} \nabla^2 T_1 -
(\gamma-1)(T_s+\frac{T_o}{\Delta T}) \nabla \cdot \bm v 
\end{equation}
which simplifies for $\frac{\Delta T}{T_o} \ll 1$ and $\frac{d}{H_o} \ll 1$ to
\begin{equation}
\partial_t T_1 - v_z = \frac{\gamma}{\rho_o} \nabla^2 T_1 -
(\gamma-1)\frac{T_o}{\Delta T} \nabla \cdot \bm v .
\label{eq:lin_temp}
\end{equation}
Finally, the linearized equation of continuity is
$\partial_t \rho_1 + \nabla \cdot (\rho_s \bm v)$, or
\begin{equation}
\partial_t \left( \frac{\rho_1}{\rho_o} \right)
=
-v_z \left( \frac{\Delta T}{T_o} - \gamma \frac{d}{H_o} \right)
- \nabla \cdot \bm v
\label{eq:lin_conti}
\end{equation}
in the limit of small $\frac{\Delta T}{T_o}$ and $\frac{d}{H_o}$.

The eqs. (\ref{eq:conti}-\ref{eq:T}) were time stepped with a finite difference
method of second order in space on a staggered grid. The time step was a third
order Runge-Kutta method. The code was implemented on GPUs with resolutions up
to 512 grid points in horizontal and 256 grid points in vertical direction.
The validation procedure was the same as in the appendix of ref. \onlinecite{Tilgne11}.

\begin{figure}
\includegraphics[width=8cm]{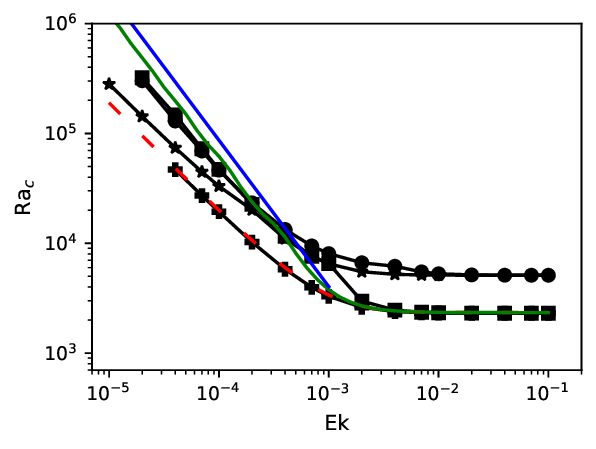}
\includegraphics[width=8cm]{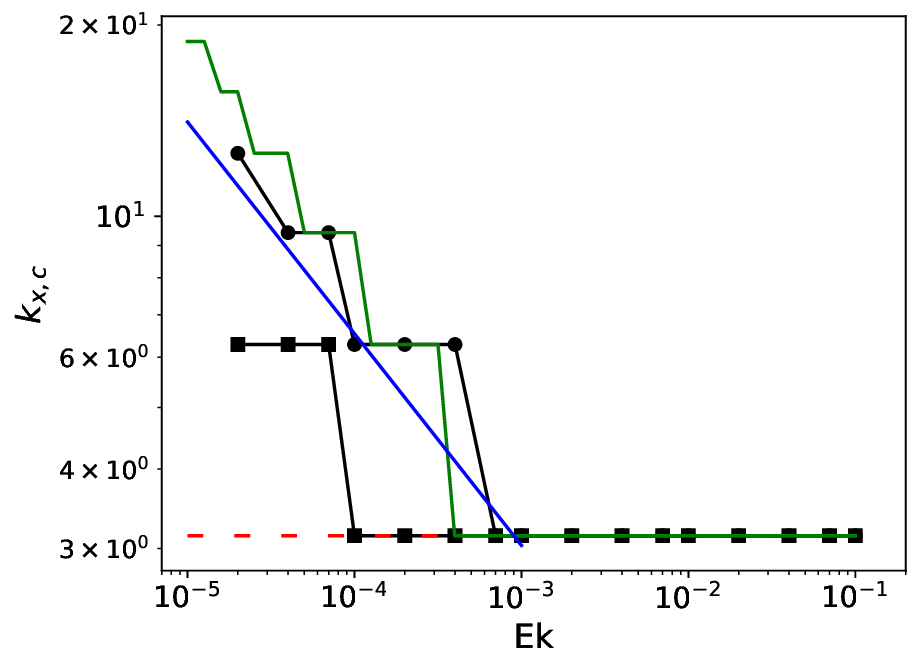}

\includegraphics[width=8cm]{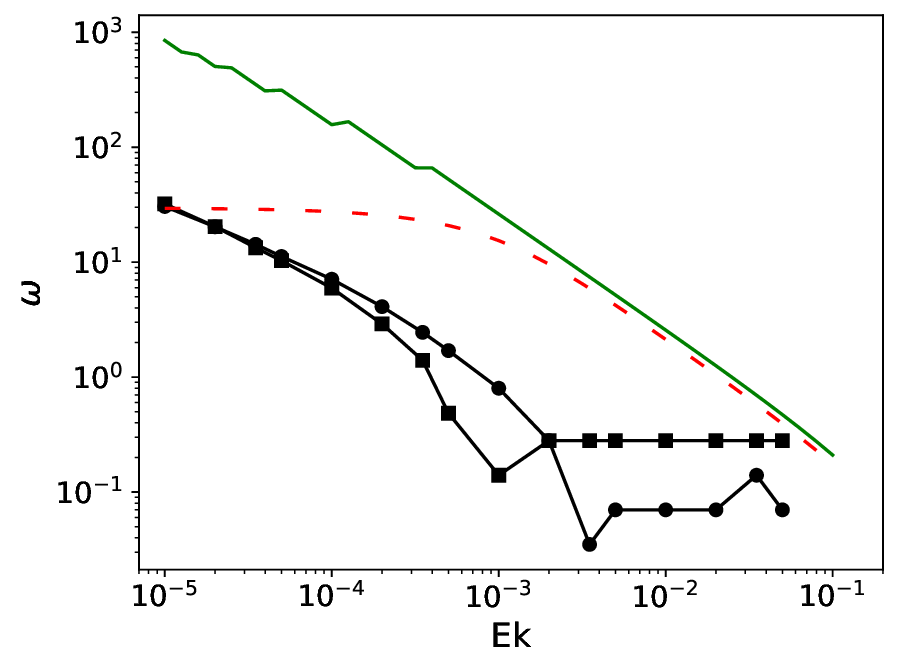}
\includegraphics[width=8cm]{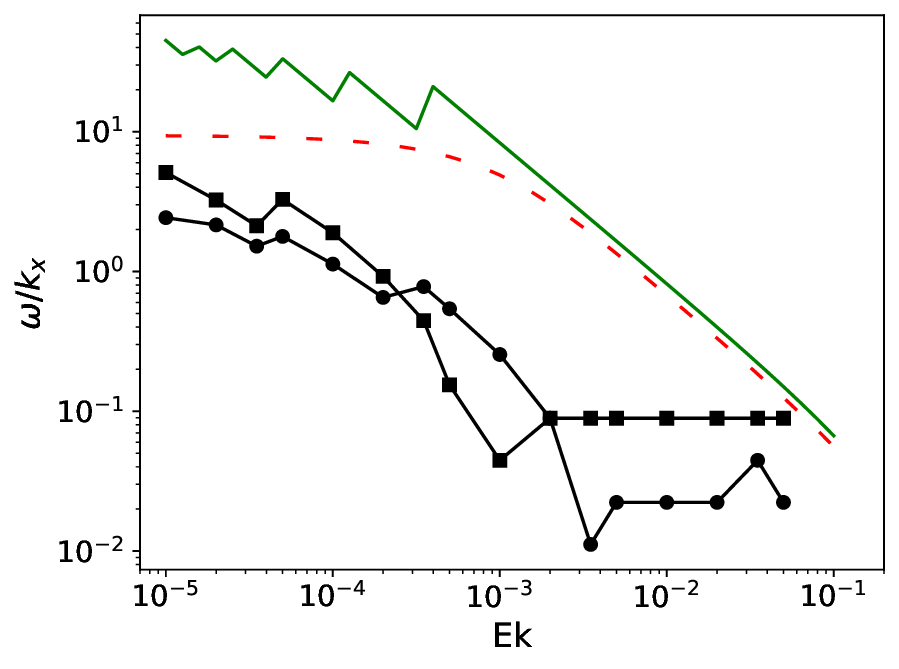}
\caption{
The critical Rayleigh number (top left), together with the wave number (top
right), the frequency (bottom left), and the phase velocity (bottom right) of
the critical mode at onset as a function of the Ekman number for $\frac{\Delta T}{T_o}=\frac{d}{H_o}=0.1$. 
The dashed red lines are the results from the linear system with constant coefficients 
(\ref{eq:lin_mom},\ref{eq:lin_temp},\ref{eq:lin_conti}) and the solid green
lines are the results from the same system with $\partial_t\rho_1$ set to
zero. The solid blue line indicates $\mathrm{Ra}_c\propto\mathrm{Ek}^{-4/3}$ in
the top left and $k_{x,c}\propto\mathrm{Ek}^{-1/3}$ in the top right panels. The
black symbols show the results from simulations of the full linearized equations
with different boundary conditions. Pluses are free slip plates with
periodic lateral boundaries, stars are also laterally periodic but have no slip
plates, circles have no slip plates with free slip lateral walls, and squares have free slip 
boundaries in all directions and have an spect ratio of $L/d=1$.}
\label{fig:onset}
\end{figure}

\section{The linear onset}
\label{section:onset}

The equations (\ref{eq:conti}-\ref{eq:T}) describe convection in a compressible
gas rotating about an axis perpendicular to gravity. This configuration allows
for vorticity generation through the compressional $\beta$-effect which is
analogous the topographic $\beta$-effect familiar from rotating spherical shells
or the $\beta$-plane. Previous studies of the compressional $\beta$-effect were
based on the anelastic approximation \cite{Evonuk08,Glatzm09,Verhoe14} in which
case the equation for the vorticity component along the rotation axis is exactly
the same for all the different instances of the $\beta$-effect. This leads to
the expectation that, all other parameters held constant, the critical Rayleigh
number should scale \cite{Busse14b} as $\mathrm{Ek}^{-4/3}$, and the wavenumber of the critical
mode should scale as $\mathrm{Ek}^{-1/3}$. 

It was also noted that there are issues with the predictions of some variants of
the anelastic approximation concerning the onset of convection which appear if
some velocity becomes comparable to the speed of sound, as for example the drift
velocity of the critical mode \cite{Verhoe18}. In rotating Rayleigh-B\'enard
convection in which rotation and gravity are aligned with each other, the
critical modes drift at onset at low Prandtl numbers. This drift can be fast
enough so that the time derivative term in the continuity equation must be
retained \cite{Calkin15}. A more comprehensive point of view was adopted by
\citet{Wood16} who investigated general linearized sound proof models and argued
that meaningful results can only be expected if the equations conserve energy.
Within this class of approximations, \citet{Wood16} identified only a single set
of equations whose dispersion relation is identical to that of the linearized
full equations, and this approximation is different from the anelastic
approximation on which most previous work is based \cite{Bragin95}.

The current state of knowledge provides us with the motivation to investigate
the onset of convection in some detail and by two different means. First of all, one
can drop all nonlinear terms from eqs. (\ref{eq:conti}-\ref{eq:T}), time step
the resulting equations and compute energy as a function of time. After
transients, the energy decays or increases depending on whether the Rayleigh
number is below or above its critical value, and the energy stays constant if the
Rayleigh number is marginal.

Since our main interest is in the weakly compressible flows realized in
experiments, we will also turn in a second approach to the simplified equations
(\ref{eq:lin_mom},\ref{eq:lin_temp},\ref{eq:lin_conti}). These equations are linear and in
addition have constant coefficients. They therefore admit solutions of the form
\begin{equation}
\frac{\rho_1}{\rho_o} = {\tilde \rho} e^{i \bm k \cdot \bm r} e^{\sigma t}
~~~,~~~
T_1 = {\tilde T_1} e^{i \bm k \cdot \bm r} e^{\sigma t}
~~~,~~~
\bm v = \bm{\tilde v} e^{i \bm k \cdot \bm r} e^{\sigma t}
\label{eq:plane_wave}
\end{equation}
where ${\tilde \rho}$, ${\tilde T_1}$ and $\bm{\tilde v}$ are complex amplitudes
independent of space and time and $\sigma$ is a complex growth rate. The linear
onset problem in 2D then reduces to the computation of the eigenvalues of a 
$4 \times 4$ matrix. We will want to take advantage of such a dramatic simplification 
and determine critical Rayleigh numbers from this purely algebraic problem.
Even though this is a much simpler problem than the
original eigenvalue problem posed by the linearized partial differential
equations, only numerical evaluations of the eigenvalues of the
$4 \times 4$ matrix are practical.

Solutions in the form of plane waves ignore the presence of boundaries. To
emulate boundaries, we restrict the admissible wavenumbers 
in eq. (\ref{eq:plane_wave}) to $k_z=\pi$ and
$k_x=n\pi$ with $n=1,2,3,...$. The solutions obtained from the matrix problem
will be compared to different sets of boundary conditions implemented in the
direct numerical simulations of the full linearized equations. Periodic boundary
conditions in the directions perpendicular to gravity accommodate plane waves at
least in this direction, so that we will solve the full linear onset problem
both with free slip and adiabatic sidewalls, and with periodic lateral boundary
conditions, but never with no slip sidewalls. Plane waves can be superposed so
that the dependence on $z$ is a sine
or cosine function which satisfies free slip boundary conditions, so that the
full eigenvalue problem will also be solved with stress free boundary conditions
for the purpose of comparison, in addition to the no slip boundary conditions
which are implemented in the nonlinear simulations of the next sections and
which are most relevant for the simulation of experiments.

Fig. \ref{fig:onset} summarizes the results for
$\frac{\Delta T}{T_o}=\frac{d}{H_o}=0.1$. These parameters are typical for a
laboratory experiment and they are small enough so that the simplifications
leading to eqs. (\ref{eq:lin_mom},\ref{eq:lin_temp},\ref{eq:lin_conti}) seem justified. The
critical Rayleigh number $\mathrm{Ra_c}$ obtained from the matrix problem agrees
remarkably well with the $\mathrm{Ra_c}$ obtained from the onset calculation for
free slip boundaries with periodic lateral boundary conditions. However, the
dependence of $\mathrm{Ra_c}$ on $\mathrm{Ek}$ in these two cases is different
from the $\mathrm{Ek}^{-4/3}$ expected from the anelastic approximation, and the
wavenumber $k_x$ of the critical mode (which is restricted to integer multiples
of $\pi$) is independent of $\mathrm{Ek}$. The computed $\mathrm{Ra_c}$ behaves
approximately as
$\mathrm{Ra_c} \propto \mathrm{Ek}^{-1}$.

To elucidate the origin of this discrepancy, the matrix problem was also solved
after omitting the term $\partial_t \rho_1$ from eq. (\ref{eq:lin_conti}). The
equation of continuity then has the same form as in the anelastic approximation.
The critical Rayleigh number obtained from this somewhat artificial problem is
also shown in fig. \ref{fig:onset}. At small Ekman numbers, one finds
$\mathrm{Ra_c} \propto \mathrm{Ek}^{-4/3}$ and $k_c \propto \mathrm{Ek}^{-1/3}$
as predicted by the anelastic approximation. When the Ekman number is large and
rotational effects are negligible, the onset of convection is stationary and the
computed $\mathrm{Ra_c}$ is of course independent of whether the term 
$\partial_t \rho_1$ was included or not. But at 
$\frac{\Delta T}{T_o}=\frac{d}{H_o}=0.1$, an $\mathrm{Ek}$ of $10^{-3}$ or less
is already enough to produce a very noticeable discrepancy between the two forms of the
continuity equation. The scalings
$\mathrm{Ra_c} \propto \mathrm{Ek}^{-4/3}$ and $k_c \propto \mathrm{Ek}^{-1/3}$
apparently require the linearized equation of continuity as it
appears in the anelastic approximation, which is
\begin{equation}
v_z \left( \frac{\Delta T}{T_o} - \gamma \frac{d}{H_o} \right)
+ \nabla \cdot \bm v = 0.
\label{eq:anelastic_conti}
\end{equation}
This equation imposes a constraint on the structure of the velocity field that
the critical mode of the full problem apparently does not satisfy. The imaginary part of
$\sigma$ still is small compared with the inverse of the travel time of sound
waves across the cell or the rotation rate of the frame of reference so that the
reasons for the failure of the anelastic approximation identified by
\citet{Verhoe18} do not apply here. But the imaginary part of $\sigma$ is large
enough so that the time derivative term needs to be retained in the equation of
continuity. Rapid oscillations in the critical mode have already been found to
be an obstacle for the anelastic approximation in rotating Rayleigh-B\'enard
convection in plane layers \cite{Calkin15} and in convection in rotating
spherical shells \cite{Liu19}.

\begin{figure}
\includegraphics[width=8cm]{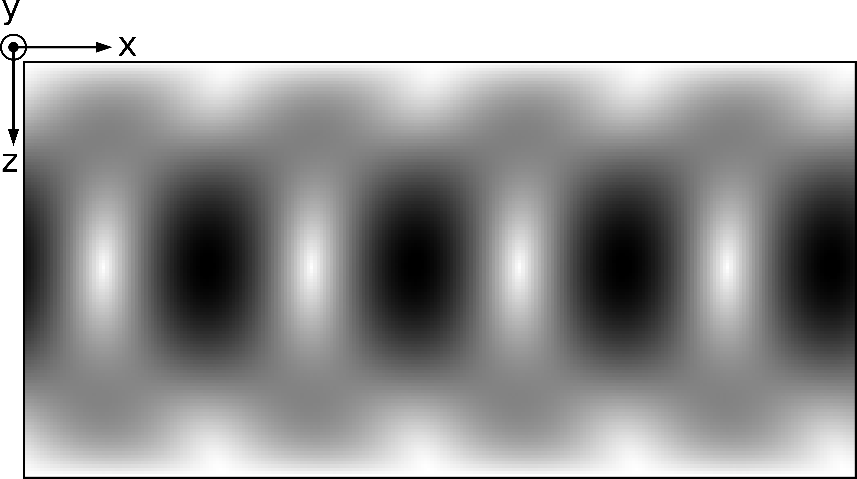}
\includegraphics[width=8cm]{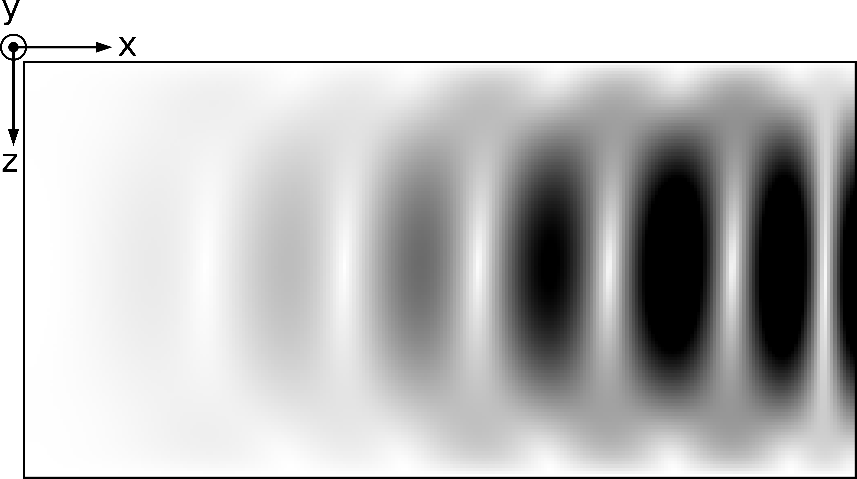}
\caption{Snapshots of the velocity magnitude $|\bm v|$ obtained from the full
linearised equations. The boundary conditions are no-slip in $z$, free slip in
$y$ and periodic (left panel) or free slip (right panel) in $x$. Both
simulations are for  $\mathrm{Ek}=4\times10^{-5}$ and $\mathrm{Ra}$ close to the
critical Rayleigh number ($\mathrm{Ra}=7.4\times10^{4}$ in the left panel  and
$\mathrm{Ra}=1.3\times10^{5}$ in the right panel). These snapshots are extracted
from animations contained in the supplementary material.}
\label{fig:beauty_shots}
\end{figure}

We now turn to the stability limits determined from the full linearized
equations. As already mentioned, periodic lateral boundary conditions with free
slip conditions in $z$ lead to convincing agreement with the algebraic problem.
The critical Rayleigh number increases if the boundary conditions become no
slip, more so with sidewalls than with periodic lateral boundary conditions.
We also performed a series of computations with free slip conditions in $z$
together with sidewalls, separated by only 1 nondimensional unit as opposed to
2 for all the other computations. The marginal mode of the periodic case still
fits into this box with stress free sidewalls. The critical mode drifts because 
of the compressional $\beta$-effect in the periodic case at small $\mathrm{Ek}$. 
Sidewalls obviously are an obstacle to the drift. Perhaps
surprisingly, the critical mode still drifts in the presence of sidewalls.
Wavecrests are continuously generated at one sidewall which grow in amplitude while
they travel to the opposite sidewall where they are annihilated. Snapshots taken from animations
included in the supplemental material and shown in fig. \ref{fig:beauty_shots} exemplify
this behavior. The sidewalls
reduce the oscillation frequency of the mode and its phase velocity, but the time dependence 
remains fast enough to invalidate the
prediction of the anelastic approximation for $\mathrm{Ra_c}$. 

The drift velocity for no slip boundaries is neither systematically larger nor
smaller than for free slip boundaries, and the oscillation frequencies seem to
converge to each other at small $\mathrm{Ek}$ for both types of boundary
conditions.

\section{Comparison of 2D and 3D convection}
\label{section:compare}

This section will compare compressible 2D and 3D convection in nonrotating
frames. We start from the
assumption that there are combinations of parameters at
which spanwise rotation forces convection to be 2D but has no additional effect
on the flow. 
Assuming such a 2D flow is realized, how does it differ from 3D 
nonrotating convection at the same $\mathrm{Ra}$, $\mathrm{Pr}$, $\frac{d}{H_o}$
and $\frac{\Delta T}{T_o}$? To answer this question, we take advantage of the
database on 3D compressible convection in ref. \onlinecite{Tilgne11} and run 2D
simulations at the exact same parameters.

The most elementary quantities to be compared are of course the kinetic energy density
$E_{\mathrm{kin}}$
\begin{equation}
E_{\mathrm{kin}}=\langle \frac{1}{V} \int \frac{1}{2} \rho \bm v^2 dV \rangle
\label{eq:Ekin}
\end{equation}
where the integration extends over the entire volume $V$ in 3D or area in 2D.
Angular brackets denote time average. The characteristic velocity is determined
by the Peclet number $\mathrm{Pe}$
\begin{equation}
\mathrm{Pe}=  \sqrt{\langle\frac{1}{V} \int \bm v^2 dV\rangle} 
\label{eq:Pe}
\end{equation}
which is called Peclet number because velocity has been made nondimensional by
the velocity scale $\kappa_o/d$. The heat injected into the fluid is the heat
traversing the warm boundary and serves to define the Nusselt number
$\mathrm{Nu}$ in
\begin{equation}
\mathrm{Nu}=- \langle \frac{1}{A} \int \partial_z T dA \rangle
\label{eq:Nu}
\end{equation}
where $A$ is the area (or length in the 2D case) of the warm boundary. The
isentropic state in the model considered here has a uniform temperature gradient
with a temperature difference $\Delta T_{\mathrm{ad}}$ between the temperature
regulated boundaries of
\begin{equation}
\frac{\Delta T_{\mathrm{ad}}}{\Delta T} = (\gamma-1)\frac{T_o}{\Delta
T}\frac{d}{H_o}
= \frac{gd}{C_p \Delta T}.
\label{eq:Delta_T_ad}
\end{equation}
The Nusselt number $\mathrm{Nu}_*$ based on the superadiabatic heat flux is thus
defined as 
\begin{equation}
\mathrm{Nu}_* = \frac{\mathrm{Nu} - \Delta T_\mathrm{ad}/\Delta T}{1-\Delta
T_\mathrm{ad}/\Delta T}.
\label{eq:Nu_NB}
\end{equation}
It is useful to similarly define a Rayleigh number $\mathrm{Ra}_*$ based on the
superadiabatic temperature difference as
\begin{equation}
\mathrm{Ra}_*=\frac{g d^3 (\Delta T- \Delta T_\mathrm{ad})}{T_o \kappa_o \nu_o}
=\mathrm{Ra} \left( 1-(\gamma-1) \frac{d}{H_o} \frac{T_o}{\Delta T} \right).
\label{eq:Ra_NB}
\end{equation}

\begin{figure}
\includegraphics[width=8cm]{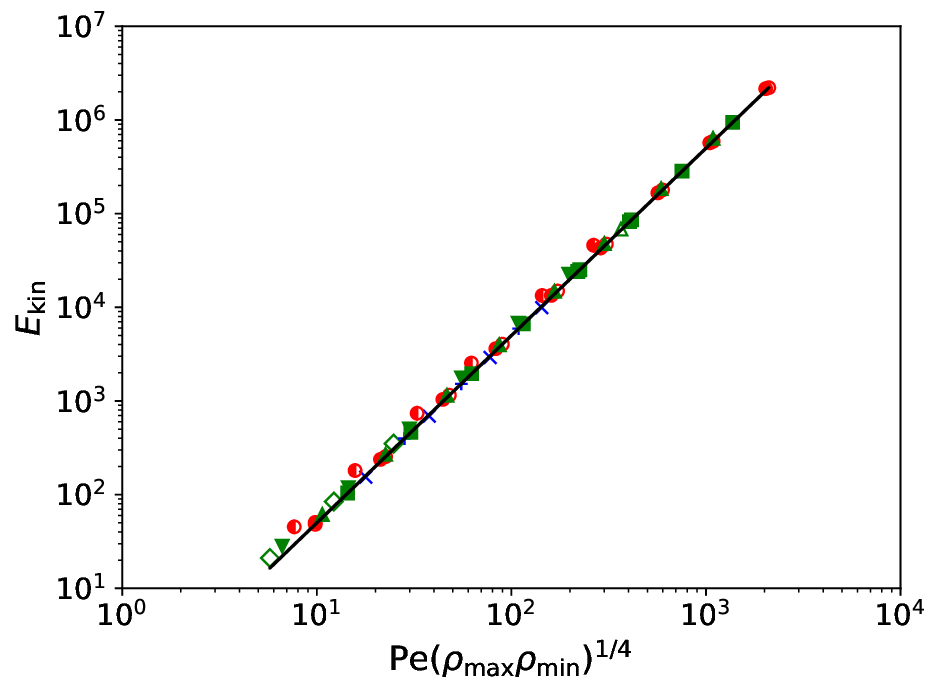}
\caption{
The kinetic energy density as a function of $\mathrm{Pe} (\rho_\mathrm{max}
\rho_\mathrm{min})^{1/4}$. The symbols are the same as in ref. \onlinecite{Tilgne11}
with $\Delta T_\mathrm{ad}/\Delta T=1/15$ (blue symbols) and $\Delta T/T_o=0.1$
(plus) or $1$ (x), $\Delta T_\mathrm{ad}/\Delta T=2/3$ (green symbols) and $\Delta T/T_o=0.1$ 
(empty squares), $0.3$ (full squares), $1$ (empty triangle up), $3$ (full triangle up), $10$ 
(empty triangle down), $30$ (full triangle down), and $100$ (empty diamonds), and $\Delta T_\mathrm{ad}/\Delta T=4/5$ 
(red symbols) and $\Delta T/T_o=0.1$ (empty circles), $1$ (full circles), and $10$ (half filled circles).}
\label{fig:Ekin_Pe}
\end{figure}

$E_{\mathrm{kin}}$ and $\mathrm{Pe}$ describe the same property in
incompressible fluids. In a compressible gas, however, the relation between 
$E_{\mathrm{kin}}$ and $\mathrm{Pe}$ depends on an effective density. It was
found for 3D convection \cite{Tilgne11} that $E_{\mathrm{kin}}$ and
$\mathrm{Pe}$ are related by
\begin{equation}
E_{\mathrm{kin}}=\frac{1}{2} \sqrt{\rho_\mathrm{max} \rho_\mathrm{min}} ~
\mathrm{Pe}^2
\end{equation}
where $\rho_\mathrm{max}$ and $\rho_\mathrm{min}$ are the maximum and minimum of
the density profiles obtained after averaging $\rho(\bm r,t)$ over time, $x$ and
$y$. The exact same relation holds in 2D (see fig. \ref{fig:Ekin_Pe}) so that
$\sqrt{\rho_\mathrm{max} \rho_\mathrm{min}}$ acts as effective density both in
3D and in 2D.

Convection in a compressible gas also differs from convection within the
Boussinesq approximation in that the two thermal boundary layers near the
temperature regulated boundaries are not symmetric to each other \cite{Jones22}. One can
compute velocities from quantities local to each boundary layer. There is a free
fall velocity, which depends on the density and temperature profiles within the
boundary layers, and there are the maximal velocities parallel to the
boundaries. There are two relations connecting these velocities in 3D convection
(figs. 8 and 9 in ref. \onlinecite{Tilgne11}) and these relations are again valid in 2D.

\begin{figure}
\includegraphics[width=8cm]{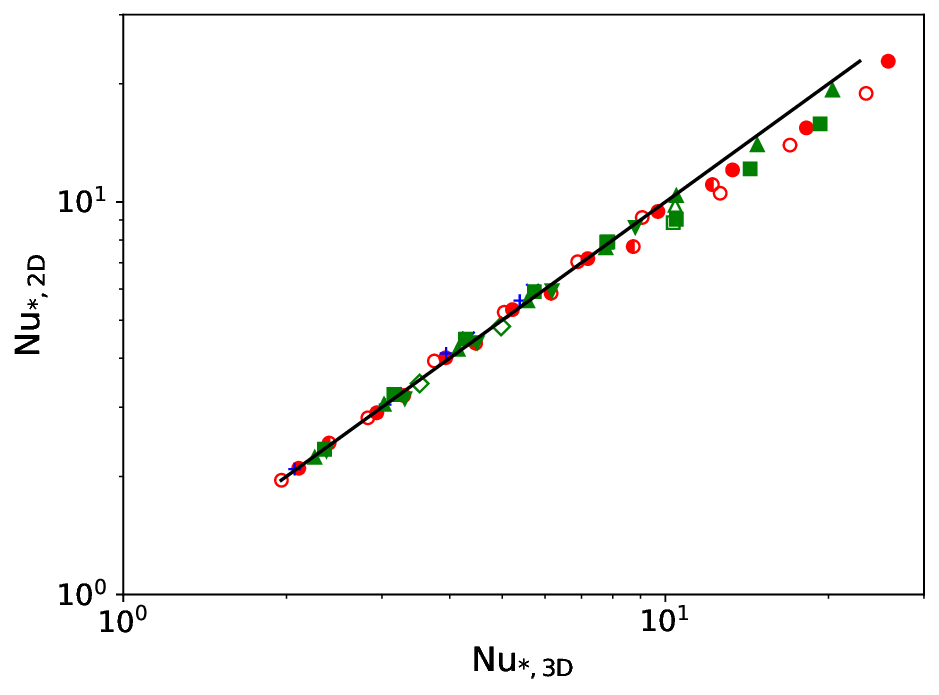}
\caption{
Comparison between $\mathrm{Nu}_*$ in the 3D simulations from ref.
\onlinecite{Tilgne11}, $\mathrm{Nu}_{*,3D}$, and the 2D simulations at the identical
control parameters, $\mathrm{Nu}_{*,2D}$. The
symbols are the same as in Figure \ref{fig:Ekin_Pe}. The black line is the
diagonal $\mathrm{Nu}_{*,3D}=\mathrm{Nu}_{*,2D}$.}
\label{fig:Nu2D_Nu3D}
\end{figure}

Other quantities of interest are detectably different in 2D and 3D. Since 2D and
3D data are available for the same control parameters, we can directly compare
the superadiabatic Nusselt number for 2D flow, $\mathrm{Nu}_{*,2D}$, with its
analog in 3D convection, $\mathrm{Nu}_{*,3D}$. As seen in fig.
\ref{fig:Nu2D_Nu3D}, the two Nusselt numbers are equal at small 
$\mathrm{Nu}_{*}$ and $\mathrm{Nu}_{*,2D}$ becomes
smaller than $\mathrm{Nu}_{*,3D}$ for large $\mathrm{Nu}_{*}$. 
However, the difference between the Nusselt numbers is always less than 20\% in fig. 
\ref{fig:Nu2D_Nu3D}.

\begin{figure}
\includegraphics[width=8cm]{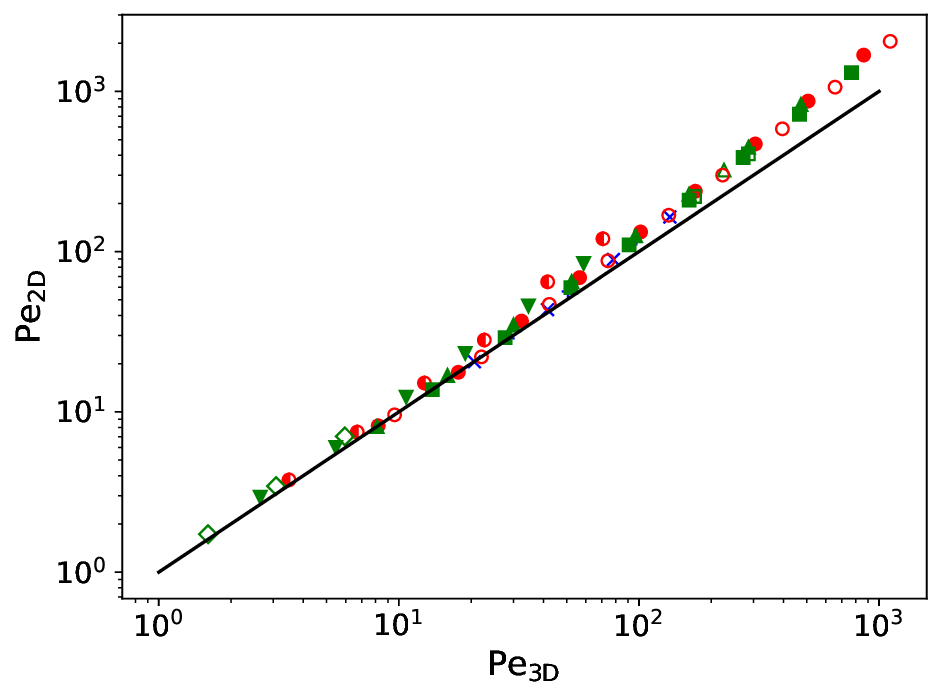}
\caption{
Comparison between $\mathrm{Pe}$ in the 3D simulations from ref. 
\onlinecite{Tilgne11}, $\mathrm{Pe}_{3D}$, and the 2D simulations at the identical 
control parameters, $\mathrm{Pe}_{2D}$. The
symbols are the same as in Figure \ref{fig:Ekin_Pe}. The black line is the
diagonal $\mathrm{Pe}_{3D}=\mathrm{Pe}_{2D}$.}
\label{fig:Pe2D_Pe3D}
\end{figure}

A considerably larger effect is visible in the Peclet number in going from its
value in 2D, $\mathrm{Pe}_{2D}$, to its magnitude in 3D, $\mathrm{Pe}_{3D}$.
Again, the two Peclet numbers are identical at small $\mathrm{Pe}$, but at
$\mathrm{Pe} \gtrsim 30$, $\mathrm{Pe}_{2D}$ exceeds $\mathrm{Pe}_{3D}$. The
marginally stable mode of the linear stability problem is two dimensional. It
therefore is not surprising that the two Peclet numbers are equal near onset
when $\mathrm{Pe}$ is small. If on the contrary the Peclet number is large, the
flow is likely prone to 3D  instabilities and the nonlinear advection term
becomes important. Nonlinearity ultimately leads to turbulence, which
notoriously behaves differently in 2D and 3D. Nonlinearity in 3D also adds a
toroidal component to the purely poloidal velocity field of the marginally
stable mode, whereas 2D flows always are purely poloidal. The increasing
fraction of toroidal velocity in 3D convection was already shown in
Boussinesq convection to lead to different global properties in 2D and 3D convection
\cite{Schmal04}. One therefore expects to see that the Peclet numbers
become different at large $\mathrm{Pe}$, but it is not obvious why
$\mathrm{Pe}_{2D}$ should become larger than $\mathrm{Pe}_{3D}$, and even less
so why simultaneously $\mathrm{Nu}_{*,2D}$ becomes smaller than
$\mathrm{Nu}_{*,3D}$.

The difference in $\mathrm{Nu}$ and $\mathrm{Pe}$ revealed by direct comparison
between 2D and 3D implies that the scaling laws relating the observables to
control parameters are different in 2D and 3D. For instance, we find
\begin{equation}
\mathrm{Nu}_{*,3D}-1 = \frac{2}{7} (E_{\mathrm{kin}} \sqrt{\rho_\mathrm{max}
\rho_\mathrm{min}})^{1/3}
\label{equ:Nu_NB_Ekin}
\end{equation}
\begin{equation}
\mathrm{Nu}_{*,2D}-1 = 0.355 (E_{\mathrm{kin}} \sqrt{\rho_\mathrm{max}
\rho_\mathrm{min}})^{0.283}
\end{equation}
and
\begin{equation}
\mathrm{Nu}_{*,3D}= 0.22 \left( \mathrm{Ra}_* (\rho_\mathrm{max}
\rho_\mathrm{min})^{1/4} \right)^{0.265}
\label{equ:Nu_NB_Ra_NB2}
\end{equation}
\begin{equation}
\mathrm{Nu}_{*,2D}= 0.275 \left( \mathrm{Ra}_* (\rho_\mathrm{max}
\rho_\mathrm{min})^{1/4} \right)^{0.25}
\label{eq:Nus_Pe}
\end{equation}

\begin{figure}
\includegraphics[width=8cm]{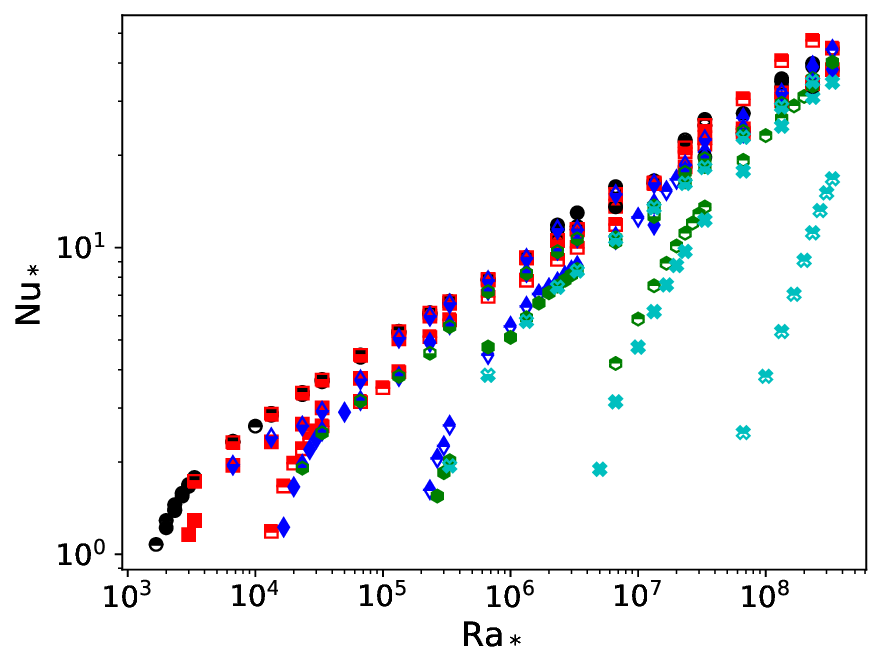}
\caption{
$\mathrm{Nu}_*$ as a function of $\mathrm{Ra}_*$ for $\Delta T/T_o=d/H_o=0.1$
(filled symbols), $\Delta T/T_o=d/H_o=0.01$ (filled in bottom half), and $\Delta
T/T_o=d/H_o=1$ (filled in top half). Symbols indicate nonrotating flows
(black circles) and the Ekman numbers $\mathrm{Ek}=10^{-3}$ (red squares), $\mathrm{Ek}=10^{-4}$ 
(blue diamonds), $\mathrm{Ek}=10^{-5}$ (green hexagons), and $\mathrm{Ek}=10^{-6}$ (cyan plus).}
\label{fig:Nu_Ra}
\end{figure}

\section{Transitions}
\label{section:transition}

The previous section omitted the Coriolis force. We will now reinstate the
Coriolis force and solve the equations (\ref{eq:conti}-\ref{eq:T}) in 2D to
determine the region in the $(\mathrm{Ra},\mathrm{Ek})-$plane in which
rotational effects are important. The
resulting Nusselt number is shown in fig. \ref{fig:Nu_Ra}. This figure is for 
$\frac{\Delta T}{T_o}=\frac{d}{H_o}=0.1$ representative of experimental
parameters as well as for $\frac{\Delta T}{T_o}=\frac{d}{H_o}=1$, and $\frac{\Delta T}{T_o}=\frac{d}{H_o}=0.01$.
One recognizes that the onset is delayed by rotation. As the
Rayleigh number is increased beyond its critical value, one first observes a
steep increase of $\mathrm{Nu}$ as a function of $\mathrm{Ra}$ until the
$\mathrm{Nu}(\mathrm{Ra})$ dependence asymptotically approaches the same
dependence as in the nonrotating case.

This figure qualitatively explains fig. 9 of \citet{Menaut19} in which one sees
a scaling in $\mathrm{Nu} \propto \mathrm{Ra}^{0.3}$ typical of nonrotating
convection at large $\mathrm{Ra}$, together with a steeper and Ekman number
dependent $\mathrm{Nu}(\mathrm{Ra})$ at low $\mathrm{Ra}$.
\citet{Menaut19} also report a hysteresis in which $\mathrm{Nu}(\mathrm{Ra})$
continues to follow the scaling of nonrotating convection if $\mathrm{Ra}$ is
lowered from large values down to the range in which the Ekman number 
dependent scaling is observed if $\mathrm{Ra}$ is increased starting from small
values. A hysteresis of this type could not be reproduced in the simulations,
possibly because the Ekman number in the simulations is orders of magnitudes
larger than in the experiments.

\begin{figure}
\includegraphics[width=8cm]{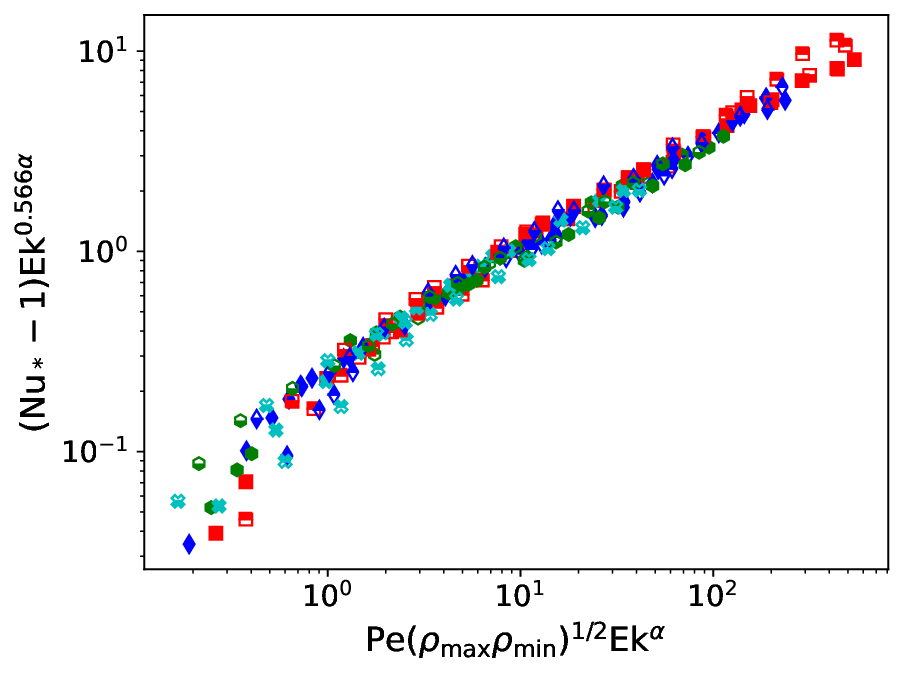}
\caption{
The same data as in Figure \ref{fig:Nu_Ra} with the same symbols represented as 
$(\mathrm{Nu}_*-1) \mathrm{Ek}^{0.566 \alpha}$ as a function of
$\mathrm{Pe} \sqrt{\rho_\mathrm{max} \rho_\mathrm{min}} \mathrm{Ek}^{\alpha}$ 
with $\alpha=0.36$. }
\label{fig:Simon_plot}
\end{figure}

\begin{figure}
\includegraphics[width=8cm]{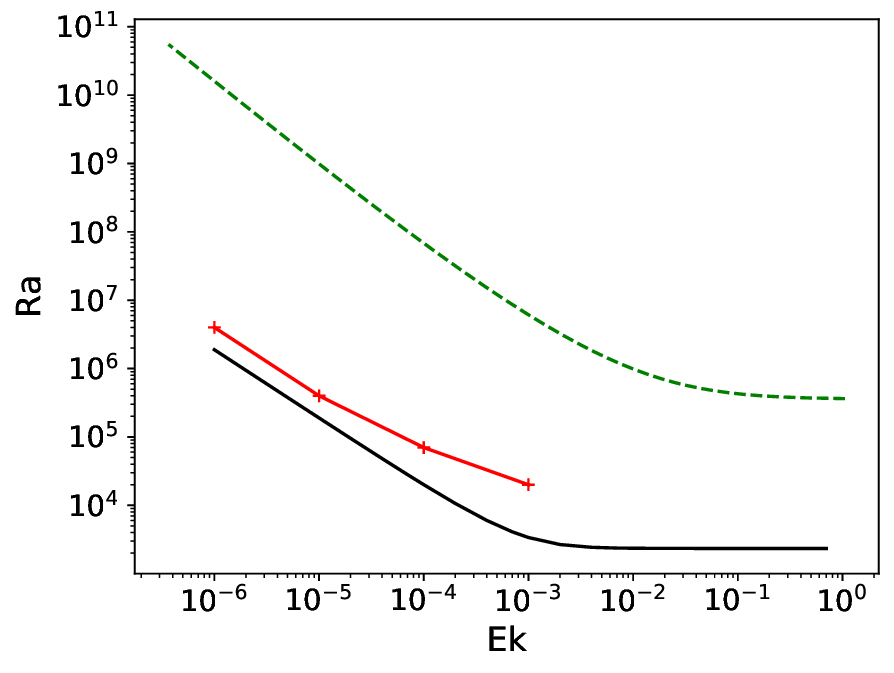}
\caption{
Shown are three transition lines in the $(\mathrm{Ek},\mathrm{Ra})-$plane with the onset of 
convection as the solid black line, the transition from 2D to 3D convection as the dashed 
green line, and the transition from rotation dependent to rotation independent scaling of 
the Nusselt number as the red pluses.}
\label{fig:Zusammenfassung}
\end{figure}

Similar transitions from flows dominated by rotation to flows nearly independent
of rotation are known from other systems. Probably the best studied example is
Rayleigh-B\'enard convection with collinear rotation and gravity axes \cite{Ecke23}.
\citet{Schmit09} presented a data reduction leading to a criterion for the
transition between both types of flows based on the Ekman and Peclet numbers.
The success of this data reduction relies on simple scaling laws valid either
near the onset of convection or in the strongly driven regime. An equally
satisfactory data collapse cannot be expected in the compressible case because
of the large number of control parameters, and because it is not trivial to
determine a scaling for the Nusselt number close to onset. Inspired by the
example of rotating Rayleigh-B\'enard convection, we represent 
$(\mathrm{Nu}_*-1) \mathrm{Ek}^{0.566 \alpha}$ as a function of 
$\mathrm{Pe} \sqrt{\rho_\mathrm{max} \rho_\mathrm{min}} \mathrm{Ek}^{\alpha}$.
The factor $0.566$ in the exponent guarantees for arbitrary $\alpha$ that the Peclet number dependence of
$\mathrm{Nu}_*$ in nonrotating convection in eq. (\ref{eq:Nus_Pe}) will appear
as a straight line in a double logarithmic plot. The exponent $\alpha$ is 
optimized to collapse the data on a single curve as well as possible. An optimum
exists near $\alpha=0.36$ which leads to fig. \ref{fig:Simon_plot}. The
nonrotating asymptote is reached for 
$\mathrm{Pe} \sqrt{\rho_\mathrm{max} \rho_\mathrm{min}} \mathrm{Ek}^{0.36}
\gtrsim 1$,
whereas the data points fan out below this transition point and the Nusselt
number differs from nonrotating convection. There is of course some uncertainty
where exactly to localize the transition so that we give the most weight to the
computations with $\frac{\Delta T}{T_o}=\frac{d}{H_o}=0.1$ which are close to
experimental conditions. Once one has defined the Peclet number at which the
transition occurs, one can use the $\mathrm{Pe}(\mathrm{Ra})$ dependence to find
the transitional Rayleigh number for every Ekman number. This procedure results
in a line in the $(\mathrm{Ek},\mathrm{Ra})-$plane which separates flows subject
to a sufficiently strong $\beta-$effect to modify the Nusselt number from flows
whose Nusselt number is indistinguishable from that of nonrotating 2D
convection.

This line is shown in fig. \ref{fig:Zusammenfassung} together with the
approximate stability limit of 2D flows. The transition from 2D to 3D flow was
not tracked by direct simulation but was instead deduced from the stability
criterion for elliptic instability. It was shown in ref. \onlinecite{Luedem22} that 2D
flows in convection cells with spanwise rotation are unstable to 3D
perturbations because of the finite ellipticity of the streamlines of the
convection rolls. The elliptical instability is inoperative if the streamlines
are either parallel or circular. There is therefore some ellipticity at which
the convection rolls are least stable. The stability limit for this ellipticity
was computed according to the method detailed in ref. \onlinecite{Luedem22} which first
yields a stability curve in the $(\mathrm{Ek},\mathrm{Pe})-$plane which must be
converted to a stability curve in the $(\mathrm{Ek},\mathrm{Ra})-$plane with the
help of the $\mathrm{Pe}(\mathrm{Ra})$ dependence. Combinations of $\mathrm{Ek}$
and $\mathrm{Ra}$ below this stability line in fig. \ref{fig:Zusammenfassung}
are certainly 2D.

Fig. \ref{fig:Zusammenfassung} contains one last stability criterion which
simply is the linear onset of convection. In summary, one finds that at small
Ekman numbers and the compressibility typical of laboratory experiments, there
are roughly four orders of magnitude in $\mathrm{Ra}$ beyond the critical
Rayleigh number for which the flow stays two dimensional, and out of these four
orders of magnitude, there is a factor of $3$ to $10$ in $\mathrm{Ra}$ for which the
heat transport is reduced from its value in nonrotating convection. There are thus at least 3 
decades in $\mathrm{Ra}$ in which the experiments can approximate 2D nonrotating 
compressible convection.

\section{Mean flow}
\label{section:mean_flow}

\begin{figure}
\includegraphics[width=8cm]{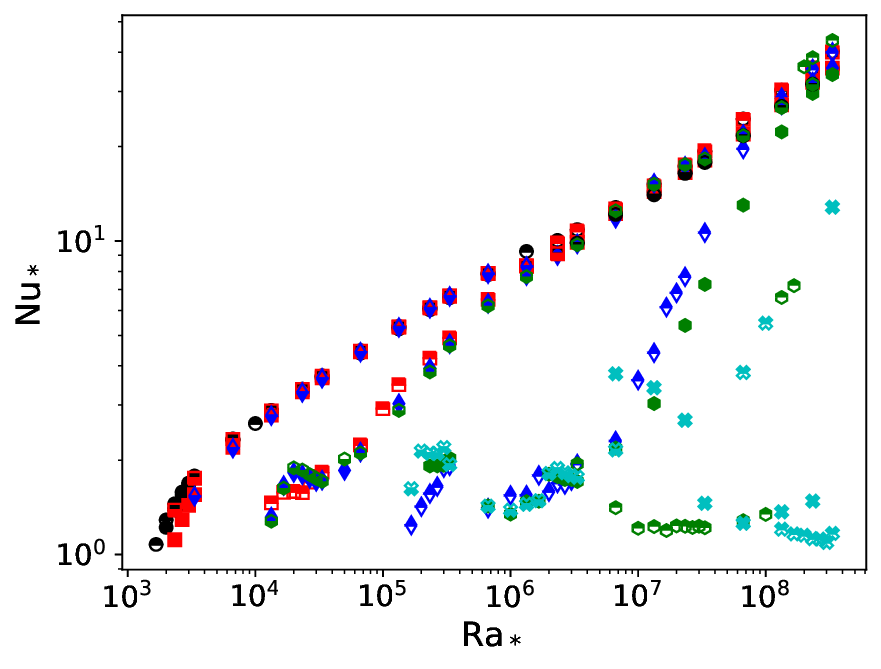}
\includegraphics[width=8cm]{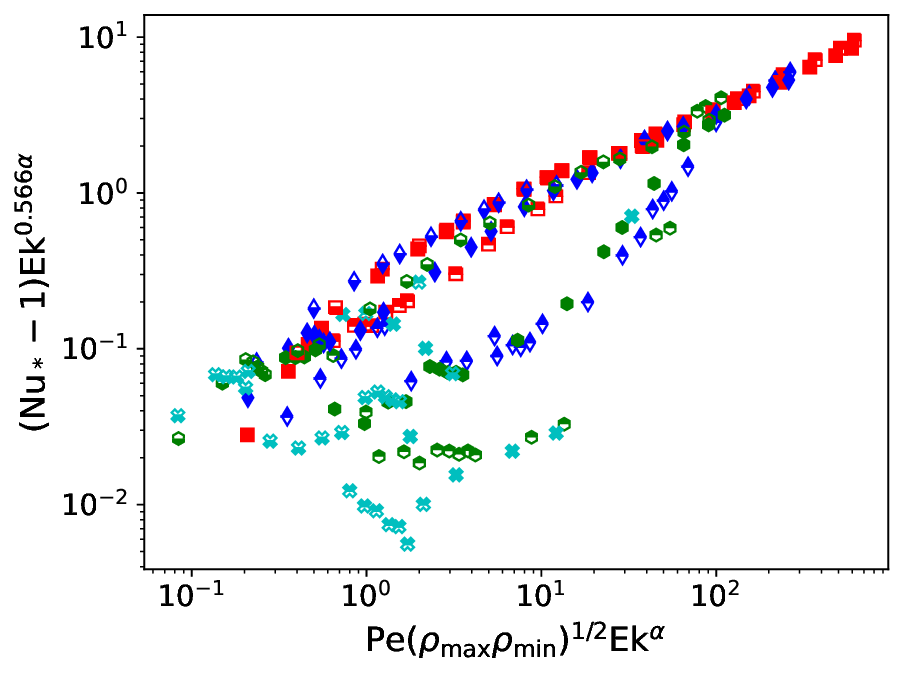}
\caption{
$\mathrm{Nu}_*$ as as function of $\mathrm{Ra}*$ (left panel) and 
$(\mathrm{Nu}_*-1) \mathrm{Ek}^{0.566 \alpha}$ as a function of
$\mathrm{Pe} \sqrt{\rho_\mathrm{max} \rho_\mathrm{min}} \mathrm{Ek}^{\alpha}$
for $\alpha=0.36$ (right panel) for the same parameters and with the same
symbols  as in Figure \ref{fig:Nu_Ra} but for periodic conditions in the
$x-$direction.}
\label{fig:mean_flow}
\end{figure}

We have seen in section \ref{section:onset} that sidewalls slow down the
drifting motion of the critical mode at onset, but not enough so to make a
qualitative difference. Another possible effect of sidewalls is the suppression
of mean flows. Experimentally, one can either use cylindrical cells
\cite{Jiang20} which allow mean flows or rectangular cells \cite{Menaut19} which
inhibit mean flows. An accurate numerical representation of a cylindrical
annulus would require a cylindrical computational volume, or at least a
Cartesian geometry with periodic lateral boundary conditions with a large aspect
ratio. However, we chose to run simulations in the Cartesian geometry with no
slip boundaries in $z$ and periodic sidewalls with an aspect ratio of only 2 
to allow for direct comparison with the results of the previous sections.

Fig. \ref{fig:mean_flow} shows the Nusselt number obtained with these boundary
conditions for 2D convection. At large $\mathrm{Ra}$, all curves approach a
common asymptote which corresponds to nonrotating convection. At $\mathrm{Ra}$
slightly above onset, the Nusselt number behaves very differently from the case
with sidewalls in fig. \ref{fig:Nu_Ra}. The heat transport is suppressed by the
presence of a mean flow. This mean flow has both a shearing component and a
general drift in prograde direction. It is known that the amplitude of mean
flows depends on the aspect ratio of the computational volume if that aspect
ratio is modest \cite{vdPoel14b}. A detailed study of the mean flow was not
undertaken for this reason. But it now is obvious that the subdivison of the
parameter space in fig. \ref{fig:Zusammenfassung} looks qualitatively the same
if we change the lateral boundary condition, except that 
the boundary between rotation dominated and 2D nonrotating behavior is displaced.
Furthermore, the dependences of $\mathrm{Nu}$ or $\mathrm{Pe}$ on
the control parameters are affected by the lateral boundary conditions.

\section{Conclusion}

The simulations presented in this paper investigate the different types of
flows that may be expected in a rectangular convection cell filled with
compressible gas placed in a centrifuge. The compressibility leads to a
compressional $\beta$-effect which causes the critical mode at onset to drift.
This drift is slowed but not removed by sidewalls and leads to a time dependence fast
enough so that the time derivative term in the continuity equation is not
negligible as would be necessary for the anelastic approximation to be
applicable. This effect must be expected to persist in different geometries of
astrophysical relevance, such as a spherical shell. 

Three distinct types of flow are accessible to convection experiments in a
centrifuge according to theory and numerical simulations: convection in
which global rotation is fast enough to force the flow to be 2D and in which the
compressional $\beta$-effect causes a drift and limits the heat flow, 2D
convection which behaves like nonrotating 2D convection in as far as the heat
flow is concerned, and 3D convection. It is not excluded that the anelastic
approximation becomes useful again once the Rayleigh number is large enough so
that the influence of the compressional $\beta$-effect on the heat flow becomes small. 

The 2D convection can only serve as an approximate
model for 3D convection since the scaling laws for heat flux and kinetic
energies are not exactly the same, even though other properties are identical.
For instance, the boundary layers near the temperature regulated boundaries are
not symmetric to each other because the density in the one is different form the
density in the other. However, velocity scales pertaining to the boundary layers
are connected by the same relationships in both dimensions.

Numerical simulations of course cannot reproduce the extreme control parameters
accessible to experiments which may lead to new surprising results in the
experiments. The existing experimental data show that there is a hysteresis in
the heat flux when the rotation rate is varied. This behavior was not seen in
the numerical simulations and the physical mechanism underlying such a
hysteresis remains to be understood.
\begin{acknowledgments}
This work was funded
by the Deutsche Forschungsgemeinschaft (DFG) under the grant Ti 243/13.
\end{acknowledgments}


%

\end{document}